\global\def\draftcontrol{0}

   \def\versionno{ causality }

\catcode`\@=11

\expandafter\ifx\csname draftcontrol\endcsname\relax\global\def\draftcontrol{0}
\fi

{\count255=\time\divide\count255 by 60
\xdef\hourmin{\number\count255}
\multiply\count255 by-60\advance\count255 by\time
\xdef\hourmin{\hourmin:\ifnum\count255<10 0\fi\the\count255}}
\def\draftdate{\number\month/\number\day/\number\year\ \ \ \hourmin }

\newcommand\makepapertitle{\par
  \begingroup
    \renewcommand\thefootnote{\@fnsymbol\c@footnote}%
    \def\@makefnmark{\rlap{\@textsuperscript{\normalfont\@thefnmark}}}%
    \long\def\@makefntext##1{\parindent 1em\noindent
            \hb@xt@1.8em{%
                \hss\@textsuperscript{\normalfont\@thefnmark}}##1}%
     \newpage
     \global\@topnum\z@   
     \@makepapertitle
     \thispagestyle{empty}\@thanks
  \endgroup
  \setcounter{footnote}{0}%
  \global\let\thanks\relax
  \global\let\makepapertitle\relax
  \global\let\@makepapertitle\relax
  \global\let\@thanks\@empty
  \global\let\@author\@empty
  \global\let\@date\@empty
  \global\let\@title\@empty
  \global\let\title\relax
  \global\let\author\relax
  \global\let\date\relax
  \global\let\and\relax
  \def\version{\let\version\@version\@gobble}
}
\def\@makepapertitle{%
  \newpage
   \ifnum\draftcontrol=1 {}
   \version\versionno
   \vskip 3em%
   \else
   \hfill\hbox to 3cm {\parbox{4cm}{\@pubnum}\hss}%
   \vskip 3em%
   \fi
   \begin{center}%
   \let \footnote \thanks
     {\LARGE {\@title}}%
     \vskip 1.5em%
     {\normalsize
       \lineskip .5em%
       \begin{tabular}[t]{c}%
         \@author
       \end{tabular}\par}%
     \vskip 1.5em%
     {\@bstract}%
     \end{center}%
     \vskip 1.5em
     \@date%
   \par
}

\gdef\@pubnum{}
\def\pubnum#1{%
  \gdef\@pubnum{#1}}

\gdef\@bstract{}
\def\Abstract#1{%
  \gdef\@bstract{%
   \parbox{\textwidth-0pc}{%
   \centerline{\bf Abstract}\penalty1000%
\kern.2cm%
\noindent
\renewcommand\baselinestretch{1.0}%
{#1}}}
}

\def\ps@paper{\let\@mkboth\@gobbletwo%
     \ifnum\draftcontrol=1
    \def\@oddfoot{\hbox to \textwidth{\tiny \versionno \hfil\tiny\draftdate}%
    \hskip -\textwidth \hbox to \textwidth{\hfil\rm\thepage\hfil}}%
     \else\def\@oddfoot{\hbox to \textwidth{\hfil\rm\thepage\hfil}}
     \fi
     \let\@evenfoot\@oddfoot
}

\def\body{\clearpage
          \pagestyle{paper}
    }

\def\@version#1{\ifnum\draftcontrol=1
\typeout{}\typeout{#1}\typeout{}
\vskip3mm\centerline{\hbox{\fbox{\normalsize{\tt DRAFT -- #1 -- }
                   {\draftdate}}}}\vskip3mm
\fi}
\let\version\@version
\long\def\eqlabel#1{\ifnum\draftcontrol=1
                    \tag@false  
                    \tag*{(\theequation) \hbox to -0.2cm{\hspace{0cm}\small{#1}\hss}}
                    \refstepcounter{equation}
                    \edef\@currentlabel{\theequation}
                    \ltx@label{#1}          
                    \else
                    \label{#1}
                    \fi
                    }
\let\st@bibitem\@bibitem
\let\st@lbibitem\@lbibitem
\ifnum\draftcontrol=1
  \def\@bibitem#1{%
    \st@bibitem{#1}\a@@label{#1}\ignorespaces}
  \def\@lbibitem[#1]#2{%
    \st@lbibitem[#1]{#2}\a@@label{#2}\ignorespaces}
  \def\a@@label#1{%
    \gdef\a@lab{\smash{\normalfont\small#1}}
    \ifvmode
      \if@inlabel
        \global\setbox\@labels\hbox{%
          \llap{\a@lab\let\a@lab\relax
                \kern\@totalleftmargin\kern\marginparsep}%
          \box\@labels}%
      \fi
    \fi}
\fi

\documentclass[12pt,letterpaper]{article}

\usepackage{amsmath,amssymb,array,calc,epsfig,rotating,bm}
\usepackage[sort]{cite}
\usepackage{graphicx}
\usepackage{psfrag,verbatim}

\ifnum\draftcontrol=1
\fi

\renewcommand\baselinestretch{1.25}
\renewcommand\section{\@startsection {section}{1}{\z@}%
                                   {-3.5ex \@plus -1ex \@minus -.2ex}%
                                   {2.3ex \@plus.2ex}%
                                   {\normalfont\large\bfseries}}
\renewcommand\subsection{\@startsection{subsection}{2}{\z@}%
                                   {-3.25ex\@plus -1ex \@minus -.2ex}%
                                   {1.5ex \@plus .2ex}%
                                   {\normalfont\normalsize\bfseries}}
\renewcommand\subsubsection{\@startsection{subsubsection}{3}{\z@}%
                                   {-3.25ex\@plus -1ex \@minus -.2ex}%
                                   {1.5ex \@plus .2ex}%
                                   {\normalfont\normalsize\it}}
\renewcommand\paragraph{\@startsection{paragraph}{4}{\z@}%
                                   {-3.25ex\@plus -1ex \@minus -.2ex}%
                                   {1.5ex \@plus .2ex}%
                                   {\normalfont\normalsize\bf}}

\numberwithin{equation}{section}

\def\revise#1       {\raisebox{-0em}{\rule{3pt}{1em}}%
                     \marginpar{\raisebox{.5em}{\vrule width3pt\
                     \vrule width0pt height 0pt depth0.5em
                     \hbox to 0cm{\hspace{0cm}{%
                     \parbox[t]{4em}{\raggedright\footnotesize{#1}}}\hss}}}}

\newcommand\nxt[1]  {\\\fnxt#1}

\def\cala         {{\cal A}}

\def\calb         {{\cal B}}
\def\calc         {{\cal C}}

\def\cali         {{\cal I}}

\def\calo         {{\cal O}}

\def\del          {\partial}

\def\Re           {{\rm Re\hskip0.1em}}
\def\Im           {{\rm Im\hskip0.1em}}

\def\sqr#1#2{{\vcenter{\vbox{\hrule height.#2pt
 \hbox{\vrule width.#2pt height#1pt \kern#1pt
 \vrule width.#2pt}\hrule height.#2pt}}}}

\def\a{\alpha}

\newcommand{\ww}{\mathfrak{w}}

\newcommand{\beq}{\begin{equation}}
\newcommand{\eeq}{\end{equation}}
\newcommand{\beqa}{\begin{eqnarray}}
\newcommand{\eeqa}{\end{eqnarray}}
\newcommand{\beqar}{\begin{eqnarray*}}
\newcommand{\eeqar}{\end{eqnarray*}}

\newcommand{\labell}[1]{\label{#1}} 
\newcommand{\eqlabell}[1]{\eqlabel{#1}} 
\newcommand{\reef}[1]{(\ref{#1})}
\renewcommand{\eqref}[1]{(\ref{#1})}

\newcommand{\ie}{{\it i.e.,}\ }

\newcommand{\mt}[1]{\textrm{\tiny #1}}

\newcommand{\lgb}{\l_\mt{GB}}

\newcommand{\eps}{\varepsilon}

\newcommand{\ka}{\bm{k}}

\def\a{\alpha}

\def\r{\rho}

\def\dd{{\delta}}
\def\l{\lambda}
\def\s{\sigma}
\def\rp{r_+}
\def\hf{f} 

\catcode`\@=12

\begin{document}


\title{\bf Causality of Holographic Hydrodynamics}
\pubnum{UWO-TH-09/11}


\author{
Alex Buchel$ ^{1,2}$   and Robert C. Myers$ ^{1}$ \\[0.4cm]
\it $ ^1$Perimeter Institute for Theoretical Physics\\
\it Waterloo, Ontario N2L 2Y5, Canada\\[.5em]
 \it $ ^2$Department of Applied Mathematics\\
 \it University of Western Ontario\\
\it London, Ontario N6A 5B7, Canada
 }

\Abstract{We study causality violation in holographic hydrodynamics
in the gauge theory/string theory correspondence, focusing on
Gauss-Bonnet gravity. The value of the Gauss-Bonnet coupling is
related to the difference between the central charges of the dual
conformal gauge theory. We show that, when this difference is
sufficiently large, causality is violated both in the second-order
truncated theory of hydrodynamics, as well as in the exact theory.
We find that the latter provides more stringent constraints, which
match precisely those appearing in the CFT analysis of Hofman and
Maldacena.}

\makepapertitle

\body

\version\versionno
\tableofcontents

\section{Introduction}
Hydrodynamics organizes the description of the macroscopic evolution
of systems in local, but not global, equilibrium in terms of a
derivative expansion. For concreteness, we will consider a
four-dimensional relativistic fluid here. In the simplest situation
(with no conserved charges), the dynamics of the hydrodynamic
fluctuations in the fluid is simply governed by conservation of the
stress-energy tensor $T^{\mu\nu}$,
\begin{equation}
\nabla_\nu T^{\mu\nu}=0\,. \labell{eoms}
\end{equation}
The stress-energy tensor includes both an equilibrium part (with
local energy density $\eps$ and pressure $P$) and a dissipative part
$\Pi^{\mu\nu}$,
\begin{equation}
T^{\mu \nu} =\eps\, u^{\mu}u^{\nu}+P \Delta^{\mu \nu} +\Pi^{\mu
\nu}\quad {\rm where}\ \ \Delta^{\mu \nu}=g^{\mu
\nu}+u^{\mu}u^{\nu}\,. \labell{2.11}
\end{equation}
Above, $u^\mu$ is the local four-velocity of the fluid with $u^\mu
u_\mu=-1$. Further, $\Pi^{\mu\nu} u_\nu = 0$. In phenomenological
hydrodynamics, the dissipative term $\Pi^{\mu\nu}$ can be
represented as an infinite series expansion in velocity gradients
(and curvatures, for a fluid in a curved background), with the
coefficients of the expansion commonly referred to as transport
coefficients. The familiar example of the Navier-Stokes equations
are obtained by truncating $\Pi^{\mu\nu}$ at linear order in this
expansion
\begin{equation}
\Pi^{\mu \nu}=\Pi_1^{\mu\nu}\left(\eta,\zeta\right)=-\eta\,
\sigma^{\mu \nu} -\zeta\, \Delta^{\mu \nu}\,\nabla\!\cdot\! u\,,
\eqlabell{2.14}
\end{equation}
where
\begin{equation}
\sigma^{\mu \nu}=2 \nabla^{\langle\mu}u^{\nu\rangle}\equiv
\Delta^{\mu \alpha} \Delta^{\nu \beta} \left(
\nabla_{\alpha}u_{\beta}+\nabla_{\beta}u_{\alpha}\right)
-\frac{2}{3}\Delta^{\mu \nu}\left(\Delta^{\alpha
\beta}\nabla_{\alpha}u_{\beta}\right)\,. \eqlabell{2.15}
\end{equation}
Notice that at this order in the hydrodynamic approximation we need
to introduce only two transport coefficients, namely the shear
$\eta$ and bulk $\zeta$ viscosities. In the following discussion, we
will be particularly interested in describing conformal fluids, in
which case we must also impose the vanishing of the trace of the
stress tensor. This restriction fixes $\zeta =0$, as well as
$P=\eps/3$ in the equilibrium contribution.

As noted above, hydrodynamics can be regarded as giving a systematic
derivative expansion. Within this framework,  it is then
straightforward to extend $\Pi^{\mu\nu}$ to the next order to
including terms of order $\nabla^2 u$ or $(\nabla u)^2$. In general,
this extension would require the introduction thirteen new transport
coefficients \cite{paul}. However, if we again restrict our
attention of conformal fluids, the second-order term $\Pi_2$ only
depends on five of these new transport coefficients \cite{brs3}.
While the interested reader can find the complete description in
\cite{brs3}, we only illustrate the extension here by showing the
first few new terms:
 \beqa
\Pi^{\mu\nu}&=&\Pi_1^{\mu\nu}(\eta)+
\Pi_2^{\mu\nu}\left(\eta,\tau_\Pi,\kappa,\l_1,\l_2,\l_3\right)
\eqlabell{pi2}\\
&=&-\eta\, \sigma^{\mu \nu}-\eta\,\tau_\Pi\left[{}^\langle
u\!\cdot\!\nabla \sigma^{\mu\nu\rangle} +\frac{1}{3}
\left(\nabla\!\cdot\! u\right)\,\sigma^{\mu\nu}\right]
+\lambda_1\,\sigma^{\langle\,\mu}{}_\alpha\,
\sigma^{\nu\rangle\alpha} +\cdots\,. \nonumber
 \eeqa
The terms controlled by $\lambda_{2,3}$ involve the vorticity while
$\kappa$ term is proportional to the spacetime curvature. Hence the
terms explicitly given above are sufficient to describe the
vorticity-free flow of a conformal fluid in a flat background
spacetime. As noted in \cite{brs3}, the first term proportional to
$\tau_\Pi$ essentially captures the second-order formalism of
M\"uller, Israel and Stewart (MIS) \cite{MIS} while the subsequent
nonlinear terms already represent an extension of their approach.
However, this linear term is sufficient to address the question of
causality within the hydrodynamic framework. It is well known that
if the dissipative contribution is truncated as in \reef{2.14}, for
any viscosity coefficients $\{\eta,\zeta\}$, there are always
linearized fluctuations for which the wave-front speed is
superluminal \cite{hl1}. The primary motivation of MIS was then to
eliminate this acausality in the hydrodynamic equations. Indeed the
MIS term is sufficient to tame the superluminal propagation with an
appropriate choice of the relaxation time $\tau_\Pi$, as we
demonstrate below \cite{mu}. However, we add that, as will become
evident, the constraints on $\tau_\Pi$ emerge from the behaviour of
modes outside the regime of validity of the second-order
hydrodynamic framework, \ie from very short wavelength modes. Hence,
one should keep in mind that these constraints do not signal any
fundamental pathologies but rather only indicate where a certain
approximate mathematical framework describing the fluid becomes
problematic. Nevertheless, a causal system of second-order
hydrodynamic equations is still required in many situations, such
as, numerical simulations \cite{simulate} which implicitly
extrapolate the hydrodynamic equations to the smallest numerical
scales, even though the physics of interest is in the long
wavelength regime.

Linearized fluctuations of the second-order truncated hydrodynamics
in conformal fluids were discussed in \cite{brs3}:
\nxt The dispersion relation of the shear channel fluctuations is
given by (see eq.~(3.27) of \cite{brs3})
\begin{equation}
-\ww^2\ \tau_\Pi T-\frac{i\ww}{2\pi}+\ka^2\ \frac{\eta}{s}=0\,,
\eqlabell{shear}
\end{equation}
where $\ww=\omega/(2\pi T)$ and $\ka = k/(2\pi T)$. Now the speed
with which a wave-front propagates out from a discontinuity in any
initial data is governed by \cite{fox}
\begin{equation}
\lim_{|\ka|\to\infty}\ \frac{\Re(\ww)}{\ka}\bigg|_{\rm
[shear]}=\sqrt{\frac{\eta}{s\ \tau_\Pi T}}\equiv v^{front}_{[\rm
shear]}\,. \eqlabell{cs}
\end{equation}
Hence causality in this channel imposes the restriction
\begin{equation}
\tau_\Pi T\ \ge \frac {\eta}{s}\,. \eqlabell{cash}
\end{equation}
\nxt The dispersion relation of the sound channel fluctuations is
given by (see eq.~(3.20) of \cite{brs3})
\begin{equation}
-\ww^3 \tau_\Pi T-\frac{i\ww^2}{2\pi}+\frac 13 \tau_\Pi T\
\ww\ka^2+\frac{4\eta}{3s}\ \ww\ka^2+\frac {i \ka^2}{6\pi}=0\,.
\eqlabell{sound}
\end{equation}
Hence
\begin{equation}
\lim_{|\ka|\to\infty}\ \frac{\Re(\ww)}{\ka}\bigg|_{\rm
[sound]}=\sqrt{\frac 13+\frac{4\eta}{3 s}\ \frac{1}{\tau_\Pi
T}}\equiv v^{front}_{[ \rm sound]}\,. \eqlabell{cso}
\end{equation}
From \eqref{cso}, causality in the sound channel imposes the
following condition
\begin{equation}
\tau_\Pi T\ \ge 2 \frac {\eta}{s}\,. \eqlabell{caso}
\end{equation}
From \reef{cs} and \reef{cso} above, we might note that both
$v^{front}_{[\rm shear]}$ and $v^{front}_{[\rm sound]}$ diverge as
$\tau_\Pi\rightarrow 0$. We may also see that the front velocity in
the sound channel is always larger than that in the shear
channel\footnote{In fact, this is a general result which extends to
nonconformal fluids as well \cite{revp}.} and hence the former
provides a more stringent constraint \eqref{caso} on the transport
coefficients of the second-order hydrodynamics. Again, we note that
as should be evident from \reef{cs} and \reef{cso}, these
restrictions arise from pushing the second-order hydrodynamic
framework beyond its natural regime of validity, \ie $|\ka|\ll 1$.
We return to this point in greater detail in section \ref{discuss}.

In principle, all the transport coefficients are determined by
parameters of the underlying microscopic theory. In practice, such
computations are prohibitively complicated as one has to {\it
derive} the effective theory of hydrodynamics for a given
microscopic system. The difficulties become even more insurmountable
for strongly coupled plasmas, as might be of interest at RHIC (or
the LHC). However, the AdS/CFT correspondence of Maldacena
\cite{juan,adscft} provides a new framework in which transport
coefficients are readily calculable at least for certain strongly
coupled gauge theories \cite{brs3,hydro2,review1}. Furthermore, it
is a context where the discussion of conformal fluids becomes
particularly relevant. With reference to the constraints above, one
has $\eta/s=1/(4\pi)$ \cite{etas} and $\tau_\Pi T=(2-\log 2)/(2\pi)$
\cite{brs3,hydro2} for $N=4$ super-Yang-Mills or any strongly
coupled four-dimensional gauge theory for which the holographic dual
is described by Einstein gravity \cite{bmps}. Hence a second-order
hydrodynamic analysis of such holographic plasmas does not suffer
from any problems with acausality.\footnote{A full analysis of
causality in these holographic plasmas was also discussed in
\cite{land}.}

In this paper we extend this analysis using a particular effective
model in the gauge theory/string theory correspondence.
Specifically, we consider a holographic model with a Gauss-Bonnet
(GB) gravity dual,
\begin{equation}
\cali=\frac{1}{2\ell_P^3}\int d^5 x\sqrt{-g}
\left[\frac{12}{L^2}+R+\frac{\lgb}{2}L^2 \left(R^2-4
R_{\mu\nu}R^{\mu\nu}+R_{\mu\nu\r\s}R^{\mu\nu\r\s}\right)\right]\,.
\eqlabell{gbl}
\end{equation}
The corresponding conformal gauge theory is distinguished by having
two distinct central charges \cite{two,twot} -- see section
\ref{discuss} for details. The effect of such curvature-squared
interactions on holographic hydrodynamics was examined in the
context of string theory in \cite{beyond}, however, only within a
perturbative framework. The GB gravity theory \reef{gbl} is
particularly well-behaved allowing the holographic analysis to be
extended to finite values of the coupling $\lgb$. In particular, the
ratio of the shear viscosity to the entropy density is found to be
\cite{vi2}
\begin{equation}
\frac{\eta}{s}=\frac{1}{4\pi}\Bigl[1-4\lgb\Bigr]\,. \eqlabell{etas}
\end{equation}
Below we compute the relaxation time of the CFT plasma dual to GB
gravity \eqref{gbl}, $\tau_\Pi =\tau_{\Pi}(\lgb)$.  We find that the
causality condition \eqref{caso} then constrains $\lgb$ both from
above and below. Note that in this case, the constraints are imposed
to avoid fundamental inconsistencies in the theory. Further, these
constraints on $\lgb$ would in turn lead to bounds on the viscosity
in GB hydrodynamics.

The analysis of the second-order hydrodynamics in GB gravity is
interesting because causality violations were already used to
produce an upper bound on the GB coupling in \cite{vi2a}. The
analysis there also examined the propagation of signals through the
dual gauge theory plasma but made no restriction to second-order
hydrodynamics. Hence we turn to the study of causality violation in
an exact analysis of the GB theory in section \ref{exact}. Following
\cite{ks}, the dispersion relation of physical fluctuations in  a
gauge theory plasma is identified with the dispersion relation of
the quasinormal modes of a black hole in a dual gravitational
description. Extending analysis of \cite{vi2a}, we study dispersion
relation of the GB BH quasinormal modes in the shear and the sound
channels. As was done for the scalar channel in \cite{vi2a}, we show
that requiring that these modes are not superluminar, \ie the phase
velocity remains less than one in the infinite momentum limit,
constraints $\lgb$. We find that combined these constraints are more
stringent than the causality constraints coming from the
second-order truncated GB hydrodynamics.

\section{Causality of second-order Gauss-Bonnet hydrodynamics}\label{second}

We are interested in determining when the second-order hydrodynamics
dual to GB gravity satisfies the causality constraint \reef{caso}.
Since the ratio of shear viscosity to entropy density is already
given in \reef{etas}, it only remains to determine the relaxation
time $\tau_\Pi$ for the dual plasma. The simplest approach to
discover $\tau_\Pi(\lgb)$ is to examine the dispersion relation of
the sound quasinormal mode of a GB black hole. Here, the field
theory considerations establish that \cite{brs3}:
\begin{equation}
\ww=c_s \ka-2\pi i\ \Gamma T\ \ka^2+\frac{4\pi^2 \Gamma
T}{c_s}\left(c_s^2\ \tau_\Pi T -\frac 12 \Gamma T\right)\
\ka^3+\calo(\ka^4)\,. \eqlabell{soundf}
\end{equation}
With
\begin{equation}
\Gamma T =\frac{2\eta}{3s}\quad{\rm and} \quad
c_s=\frac{1}{\sqrt{3}}\,, \eqlabell{gt}
\end{equation}
this expression is simply the Taylor series solution of the
dispersion relation \reef{sound}.

In the dual gravitational description the dispersion relation
\eqref{soundf} is obtained by imposing the incoming wave boundary
condition at the horizon and the Dirichlet condition at the boundary
on the sound channel quasinormal mode wavefunction. The technique is
clearly explained in \cite{ks}. Here, we only present the salient
steps in our analysis\footnote{Further computational details are
available from the authors upon request.}.

The planar black hole solution in GB gravity can be written as
\cite{gbbh,gbbh2}
\begin{equation}
ds^2=\frac{\rp^2}{u\,L^2}\left(-\hf(u) \cala^2 dt^2+\sum_{i=1}^3
dx_i^2\right)+\frac{L^2}{\hf(u)} \frac{du^2}{4u^2}\,,\eqlabell{bhs}
\end{equation}
where
\begin{equation}
\hf(u)=\frac{1-\sqrt{1-4\lgb(1- u^2)}}{2\lgb}\,, \eqlabell{deff}
\end{equation}
and
\begin{equation}
\cala^2=\frac 12\left(1+\sqrt{1-4\lgb}\right)\,. \eqlabell{defa}
\end{equation}
The horizon is located at $u=1$ and asymptotic boundary is reached
with $u\rightarrow 0$.\footnote{A more conventional radial
coordinate \cite{vi2,gbbh} would be given by $r^2=\rp^2/u$.
Implicitly we have also chosen the branch of well-behaved solutions
and we are restricting our considerations to $\lgb<1/4$ -- see
\cite{gbbh,gbbh2} for details.} Note that the normalization constant
$\cala$ is chosen so that $\cala^2\,f(u=0)= 1$. Hence the asymptotic
behaviour of the metric shows that the AdS curvature scale is
$\cala\,L$. The Hawking temperature, entropy density, and energy
density of the black hole are
\begin{equation}
T=\cala\ \frac{\rp}{\pi L^2}\,,\qquad
s=\frac{1}{4G_N}\left(\frac{\rp}{L}\right)^3\,,\qquad \eps=\frac 34
Ts\,. \eqlabell{thermo}
\end{equation}

Now the sound channel quasinormal mode satisfies the following
equation
\begin{equation}
Z_{[\rm sound]}''(u)+\calc_{sound}^{(1)}\ Z_{[\rm
sound]}'(u)+\calc_{sound}^{(2)}\ Z_{[\rm sound]}(u)=0\,,
\eqlabell{qsound}
\end{equation}
where the coefficients $\calc_{sound}^{(i)}$ are presented in
Appendix \ref{appa}. In the hydrodynamic limit, $\ka$, $\ww\to 0$
with $\frac{\ww}{\ka}$ kept fixed, the incoming wave boundary
condition at the horizon implies
\begin{equation}
Z_{[\rm sound]}=(1-u^2)^{-\frac{i\ww}{2}}\biggl(z_0(u;\ww,\ka)+i\ka\
z_1(u;\ww,\ka)+\ka^2\ z_2(u;\ww,\ka)+\calo(\ka^3)\biggr)\,,
\eqlabell{hydrosound}
\end{equation}
with
\begin{equation}
\begin{split}
&z_i(u;\mu\ \ww,\mu\ \ka)=z_i(u;\ww, \ka)\,,\qquad {\rm for\ any\ } \mu\,,\qquad  i=0,1,2\\
&\lim_{u\to 1}z_i(u;\ww,\ka)=\dd^0_i\,.
\end{split}
\eqlabell{conds}
\end{equation}
In \eqref{hydrosound} we kept terms in the hydrodynamic expansion to
the order necessary to identify \eqref{soundf}. The sound wave
dispersion relation \eqref{soundf} is then obtained by imposing the
Dirichlet condition on $Z_{[\rm sound]}$ at the boundary
\begin{equation}
\lim_{u\to 0} Z_{\rm sound}=0\,. \eqlabell{bc}
\end{equation}

To leading order in the hydrodynamic approximation we find
\begin{equation}
z_0=\frac{\ka^2 (x+4 \lgb-1) \cala^2-6 \lgb \ww^2 x}{2\lgb (2
\cala^2 \ka^2-3 \ww^2) x}\,, \eqlabell{z0}
\end{equation}
where we used a more convenient radial coordinate
\begin{equation}
x=\left(1-4 \lgb+4 \lgb u^2\right)^{1/2}\,. \eqlabell{xdef}
\end{equation}
Imposing the Dirichlet condition \eqref{bc} at this order recovers
the conformal sound speed: $c_s={1}/{\sqrt{3}}$.

To order $\calo(\ka)$ in \reef{hydrosound}, we find
\begin{equation}
\begin{split}
z_1=&\frac{\ww}{8\ka}\biggl( \lgb  (2 \cala^2 \ka^2-3\ww^2)\
x\biggr)^{-1}\times \biggl(2
(\cala^2\ka^2 (x+4 \lgb-1)-6\ww^2 \lgb x)\ \ln\frac{1+x}{2} \\
&+(1-x) \cala^2\ka^2 (x^2-6 x+36 \lgb x+12 \lgb-3)+6\ww^2\lgb
x(x+3)(x-1)\biggr)\,.
\end{split}
\eqlabell{z1}
\end{equation}
Imposing the Dirichlet condition \eqref{bc} at  order $\calo(\ka^2)$ identifies
\begin{equation}
\frac{\eta}{s}=\frac{1}{4\pi}\Bigl[1-4\lgb\Bigr]\,, \eqlabell{etass}
\end{equation}
in precise agreement with the result \reef{etas} which was
originally determined with a Kubo formula computation in \cite{vi2}.
\begin{figure}[t]
\begin{center}
\psfrag{beta}{{$\lgb$}} \psfrag{m}{{$\l_{min}$}}
\psfrag{p}{{$\l_{max}$}} \psfrag{caus}{{$\tau_{\Pi}
T-2\frac{\eta}{s}$}}
  \includegraphics[width=5in]{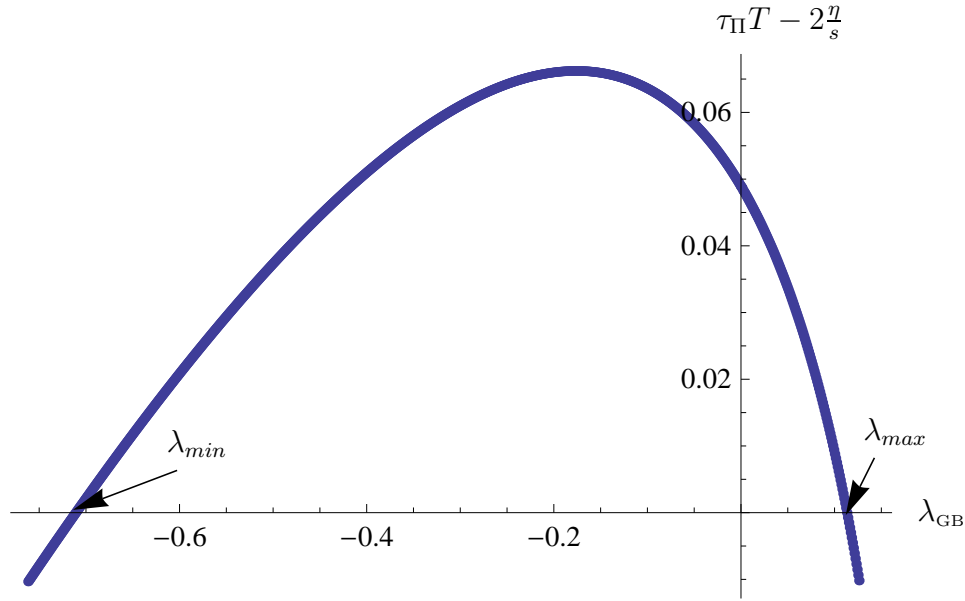}
\end{center}
  \caption{ (Colour online)
Causality of the second-order Gauss-Bonnet hydrodynamics is violated
once $\tau_\Pi T <2\frac \eta s$. Thus, $\lgb\in
[\l_{min},\l_{max}]$, where $\l_{min}=-0.711(2)$ and
$\l_{max}=0.113(0)$. }\label{fig1}
\end{figure}

\begin{figure}[t]
\begin{center}
\psfrag{vs}{{$v^{front}$}}
 \psfrag{lgb}{{$\lgb$}}
  \includegraphics[width=5in]{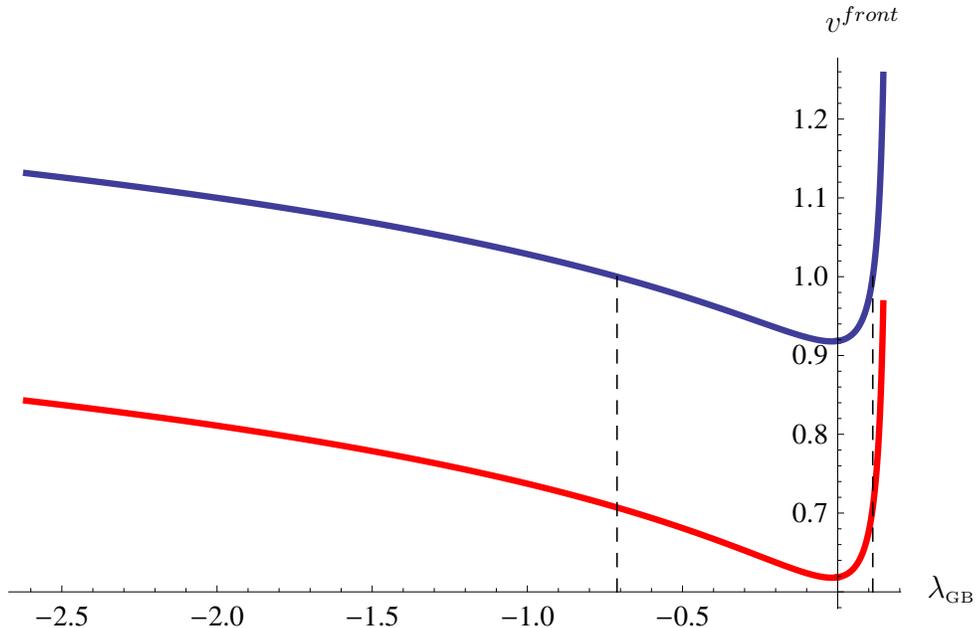}
\end{center}
  \caption{ (Colour online)
Front velocity for the shear (red) and sound (blue) channels for the
second-order hydrodynamics, as given in \reef{cs} and \reef{cso}.
The dashed vertical lines indicate $\l_{min}$ and $\l_{max}$, where
$v^{front}_{[\rm sound]}$ reaches one.}\label{fig2}
\end{figure}

Unfortunately, we were not able to evaluate $z_2$ (and as a result
$\tau_\Pi$) analytically. Thus, we had to resort to numerical
analysis. For the question of causality, we are not interested here
in $\tau_\Pi$ per se, but rather in the relation \eqref{caso}. Hence
figure~\ref{fig1} presents the difference $\left(\tau_\Pi T-2 \frac
\eta s\right)$ as a function of $\lgb$. We find that unless $\lgb\in
[\l_{min},\l_{max}]$, where $\l_{min}=-0.711(2)$ and
$\l_{max}=0.113(0)$, causality of the second-order truncated
hydrodynamics of the GB plasma is violated. In figure~\ref{fig2}, we
also present the front velocities in the shear \reef{cs} and sound
\reef{cso} channels. Note that we find that the relaxation time
vanishes for $\lgb=0.165(5)$, which causes both of the front
velocities to diverge at this point in the figure. The graph also
demonstrates that $v_{[shear]}^{front}<v_{[sound]}^{front}$, as
noted above.

\section{Causality of full Gauss-Bonnet theory}\label{exact}

In the previous section we showed that depending on the value of a
the GB coupling, $\lgb$, or equivalently on the values of the
microscopic parameters in the dual CFT, causality of the
second-order truncated hydrodynamics of the dual theory can be lost.
In this section, we wish to compare those results to causality
violations found with the analysis of \cite{vi2a}. While the latter
also looks for the appearance of superluminal signals propagating in
the dual plasma, it makes no reference to truncating the derivative
expansion in the hydrodynamic framework. We will find that the
constraints on $\lgb$ arising from the exact analysis are more
restrictive than those found in the truncated second-order
hydrodynamic analysis in the previous section.

Dispersion relation of the linearized fluctuations in plasma is
identified with the dispersion relation of the quasinormal modes of
a black hole in the dual gravitational description. There are three
types of quasinormal modes in gravitational geometries with
translationary invariant horizons \cite{ks,first}:
\nxt a scalar channel
(helicity-two graviton polarizations);
\nxt a shear channel
(helicity-one graviton polarizations);
\nxt a sound channel (helicity-zero graviton polarizations).\\
\noindent While the shear and sound channels correspond to those
considered in the previous discussion of second-order hydrodynamics,
the scalar channel was not mentioned there because it contains no
modes whose frequency vanishes as $\ka\to0$. Of course, this is in
agreement with the standard hydrodynamic analysis \cite{hl1}.
However, the scalar channel quasinormal modes of the GB black holes
in the limit $\ka\to \infty$ were studied in detail in \cite{vi2a}.
It was found there that requiring
\begin{equation}
\lim_{\ka\to \infty} \frac{\ww}{\ka}\bigg|_{[\rm scalar]}\le 1\,,
\eqlabell{zscalar}
\end{equation}
constraints $\lgb$ as follows
\begin{equation}
\lgb\le \lgb^{scalar}= \frac{9}{100}\,. \eqlabell{lscalar}
\end{equation}
Note that $\lgb^{scalar}<\l_{max}$ found in the context of the
second-order truncated hydrodynamics. In the remainder of this
section we extend analysis of \cite{vi2, vi2a} to the shear and the
sound channel quasinormal modes.

\subsection{Causality in the shear channel}

It is straightforward to derive the shear channel quasinormal
equation for the GB black holes:
\begin{equation}
Z_{[\rm shear]}''(u)+\calc_{shear}^{(1)}\ Z_{[\rm
shear]}'(u)+\calc_{shear}^{(2)}\ Z_{[\rm shear]}(u)=0\,,
\eqlabell{qshear}
\end{equation}
where the coefficients $\calc_{shear}^{(i)}$ are presented in
Appendix \ref{appb}. Following \cite{vi2a}, we now caste
\reef{qshear} into the form of the Schr\"odinger equation. Towards
this end, we introduce a new radial coordinate $y$
\begin{equation}
\frac{dy}{du}=-\frac{1}{\ u^{1/2}\hf(u)}\,, \eqlabell{dydu}
\end{equation}
and rescale the radial profile as
\begin{equation}
Z_{[\rm shear]}=\frac{1}{\calb}\ \psi_{[\rm shear]}\,,
\eqlabell{ressh}
\end{equation}
with
\begin{equation}
\begin{split}
&\frac{d}{du}\ln \calb=\biggl(4 u (2 \lgb \hf -1)^2 (\a^2 (2 \lgb
\hf -1)^2+\cala^2 \hf  (4 \lgb-1))
\biggr)^{-1}\\
&\times \biggl(\cala^2 (1-4 \lgb) (12 \hf^3  \lgb^2-16 \hf^2
\lgb-4+7 \hf )-(20 \hf^2 \lgb^2
-20 \lgb \hf \\
&+8 \lgb+3) (2 \lgb \hf -1)^2 \a^2\biggr)\,,
\end{split}
\eqlabell{db}
\end{equation}
where $\a=\ww/\ka$. The quasinormal equation \eqref{qshear} can then
be rewritten as
\begin{equation}
\begin{split}
&-\hbar^2\ \del_y^2\, \psi_{[\rm shear]} +U_{[\rm shear]}\
\psi_{[\rm shear]}
=\a^2\ \psi_{[\rm shear]}\,,\qquad \hbar\equiv \frac {1}{\ka}\,,\\
&\qquad{\rm where}\ \ \ U_{[\rm shear]}=U^0_{[\rm shear]}+\hbar^2\
U^1_{[\rm shear]}\,.
\end{split}
\eqlabell{sshear}
\end{equation}
The first part of the effective potential has the simple form when
expressed in terms of $u$
 \beqa
U^0_{[\rm shear]}(u)&=&\frac{\hf \cala^2  (1-4\lgb)}{(2\lgb\hf
-1)^2}
\eqlabell{u0sh}\\
&=& \frac{(1-4\lgb)\,(1-\sqrt{1-4\lgb(1-u^2)})}{
(1-4\lgb(1-u^2))\,(1-\sqrt{1-4\lgb})}\,, \nonumber
 \eeqa
while the expression for $U^1_{shear}$ is too long to be presented
here, but we note that the latter is a function only of $u$, $\lgb$
and $\alpha$. What is important is that in the limit $\ka\to \infty$
(or $\hbar\to0$), everywhere except in the tiny region $y\gtrsim
-\frac{1}{\ka}$ the dominant contribution to $U_{shear}$ comes from
$U^0_{shear}$. Thus in this limit we simply replace
\begin{equation}
\hbar^2\, U^1_{[\rm shear]}=\begin{cases} 0 & \text{$y<0$\,,}
\\
+\infty &\text{$y\ge0$\,.}
\end{cases}
\eqlabell{u1shear}
\end{equation}
\begin{figure}[t]
\begin{center}
\psfrag{v}{{$U^0$}}
  \includegraphics[width=5in]{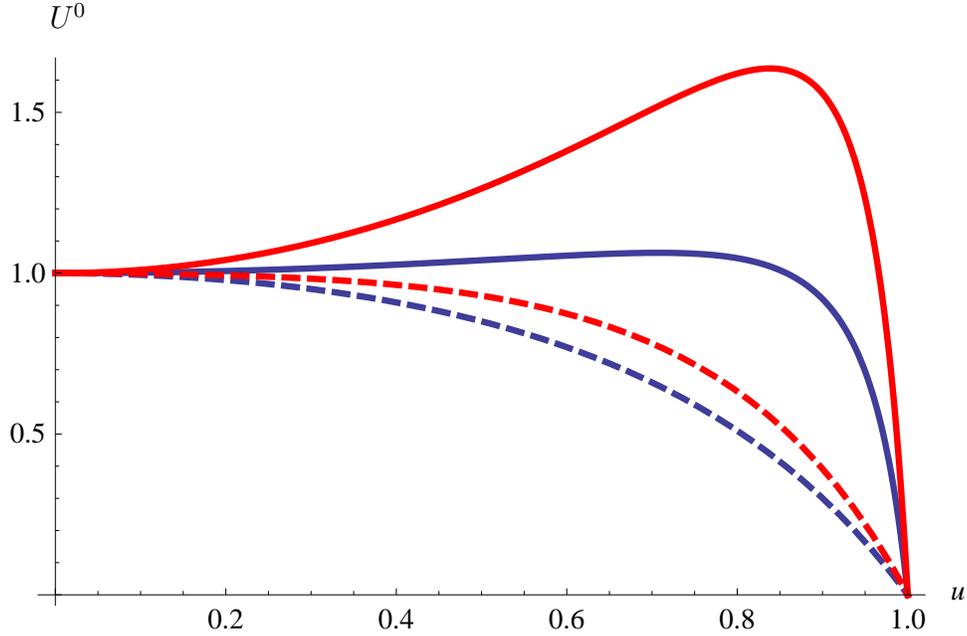}
\end{center}
  \caption{ (Colour online)
Typical behaviour of $U^0$, the leading contribution to the
Schr\"odinger potential, for both the shear (blue) and sound (red)
channels. The solid and dashed curves show the behaviour for large
and small $|\lgb|$, respectively. (Our representative values here
are: $\lgb=-1.5$ and $-0.15$.)}\label{potential}
\end{figure}

Figure \ref{potential} illustrates the general behaviour of the
leading potential \reef{u0sh}. For any values of $\lgb$, we have in
the asymptotic region, $U^0_{[\rm shear]}(u=0)=1$ while at the
horizon, $U^0_{[\rm shear]}(u=1)=0$. Now for small values of
$|\lgb|$, $U^0_{[\rm shear]}$ is a monotonically decreasing function
between these two points. However, for larger negative values of
$\lgb$, the potential develops a (single) maximum at intermediate
value of $u$:
 \beq
 U^0_{max} = \frac{1-4\lgb}{4(\sqrt{1-4\lgb}-1)}\qquad{\rm at}
 \ \ \ u_{max} = -\frac{\sqrt{\lgb\,(3+4\lgb)}}{2\lgb}\,.
 \labell{maxshear}
 \eeq
As might be inferred from $u_{max}$ above, the critical coupling for
the appearance of this maximum is $\lgb=-3/4$. At this stage, the
analysis is identical to that for the scalar channel studied in
\cite{vi2a}. Once the effective potential in the Schr\"odinger
problem \eqref{sshear} develops this new maximum, there always exist
quasinormal modes with $\Re\left(\a^2\right)\simeq U^0_{max}>1$.
This implies then that in the limit of infinite $\ka$,
$\Re(\ww)/\ka>1$ for these modes and hence they lead to a violation
of causality. Hence requiring the excitations in the shear channel
to be well behaved imposes the constraint:
\begin{equation}
\lgb\ge \lgb^{shear}=-\frac 34\,. \labell{critlgb}
\end{equation}

\subsection{Causality in the sound channel}
The quasinormal equation for the sound channel is given in
\eqref{qsound}. Following \cite{vi2a} (as reviewed in the previous
section), we arrive at the corresponding Schr\"odinger problem in
the sound channel
\begin{equation}
\begin{split}
&-\hbar^2\ \del_y^2\, \psi_{[\rm sound]} +U_{[\rm sound]}\
\psi_{[\rm sound]}=\a^2\ \psi_{[\rm sound]}\,.
\end{split}
\eqlabell{ssound}
\end{equation}
Once again, in the limit $\ka\to \infty$, potential  $U_{[\rm
sound]}$ is given by
\begin{equation}
U_{[\rm sound]}=\begin{cases} U^0_{[\rm sound]} & \text{$y<0$\,,}
\\
+\infty &\text{$y=0$\,.}
\end{cases}
\eqlabell{usound}
\end{equation}
where
 \beqa
U^0_{[\rm sound]}&=&\frac{(1-8 \lgb-4 \hf \lgb  (\lgb \hf -1)) A^2
\hf}{(2 \lgb \hf -1)^2} \eqlabell{u0sound}\\
&=& \frac{(1-4\lgb(1+u^2))\,(1-\sqrt{1-4\lgb(1-u^2)})}{
(1-4\lgb(1-u^2))\,(1-\sqrt{1-4\lgb})}\,. \nonumber
 \eeqa
The general behaviour of this potential is the same as described in
the previous section, as can be seen in figure \ref{potential}. In
particular, to avoid the appearance of an intermediate maximum in
the potential \eqref{u0sound} and the corresponding
causality-violating quasinormal modes, we must impose the constraint
\begin{equation}
\lgb \ge \lgb^{sound}=-\frac{7}{36}\,. \eqlabell{lsound}
\end{equation}
Note that $\lgb^{sound}>\lgb^{shear}$ found above and also
$\lgb^{sound}>\l_{min}$ found in the context of the second-order
truncated hydrodynamics.

\section{Conclusion}\label{discuss}
In this paper, we have shown how causality of the near-equilibrium
phenomena in the quantum field theory can be used to distinguish
``healthy'' models from the ``sick'' ones. To illustrate the point,
we studied the fluctuations in the CFT plasmas, holographically dual
to Gauss-Bonnet gravity \reef{gbl}. Our analysis found two sets of
constraints on the GB coupling:
 \beqa
\text{second-order hydrodynamics:} &
-0.711\le\lgb\le0.113\ ,&\eqlabell{results1}\\
\text{exact analysis:} & -\frac{7}{36}\le\lgb\le \frac{9}{100}\ .&
 \eqlabell{results2}
 \eeqa
It is clear that the exact analysis of section \ref{exact} produced
more stringent restrictions than the analysis of the truncated
second-order hydrodynamic equations in section \ref{second}. While
both of these approaches are examining the behaviour of
gravitational fluctuations in the GB black hole background
\reef{bhs}, the relevant quasinormal modes are very different in the
two cases. The second-order hydrodynamics is focused entirely on
the behaviour of the sound mode, \ie the lowest quasinormal mode in
the sound channel. In contrast, the potential causality violation by
highly excited quasinormal modes in the scalar channel set the upper
bound in \reef{results2} while the lower bound arises from a similar
set of quasinormal modes in the sound channel.

We must emphasize that the status of these constraints differs at a
very basic level. Theories outside of the bounds given in
\reef{results2} are fundamentally pathological. In contrast, the
constraints \reef{results1} simply indicate where a certain
approximate description of the fluid becomes problematic. As such,
it is somewhat remarkable then that the bounds coming from these two
very different approaches seem to be fairly close to each other. It
is also satisfying that the fundamental constraints \reef{results2}
are the most restrictive so that the truncated second-order
hydrodynamics will be stable in any of the cases where the
underlying theory is physically sound at a fundamental level. A
priori, this does not seem to be required by any basic principles.

\begin{figure}[t]
\begin{center}
\psfrag{k}{{$\ka$}} \psfrag{w}{{$\Re(\ww)/\ka$}}
  \includegraphics[width=5in]{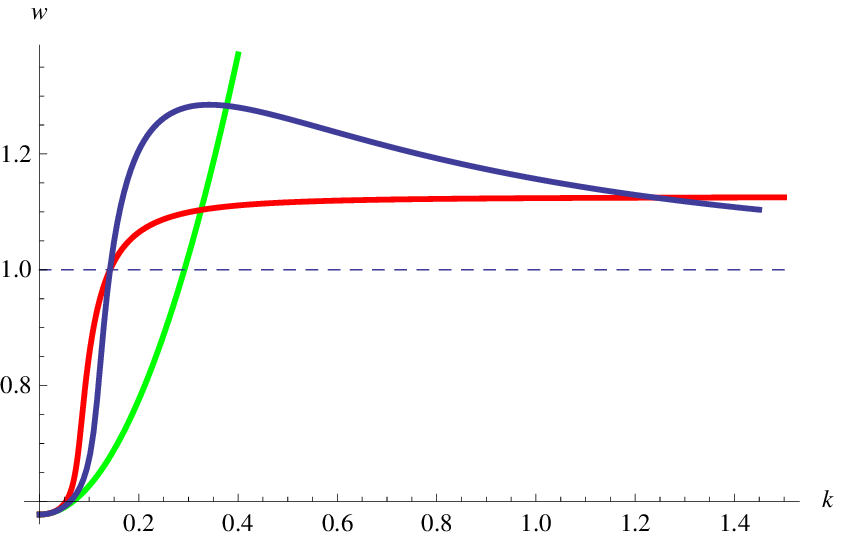}
\end{center}
  \caption{ (Colour online)
The phase velocity Re$(\ww)/\ka$ for $\lgb=-2.5$: The blue line
shows the behaviour of the lowest quasinormal mode calculated
numerically. The red curve corresponds the second-order hydrodynamic
approximation \reef{sound}. The green curve corresponds the next
order Taylor expansion \reef{soundf} arising from the second-order
dispersion relation.}\label{fig4}
\end{figure}

\begin{figure}[t]
\begin{center}
\psfrag{k}{{$\ka$}} \psfrag{w}{{$\Im(\ww)$}}
  \includegraphics[width=5in]{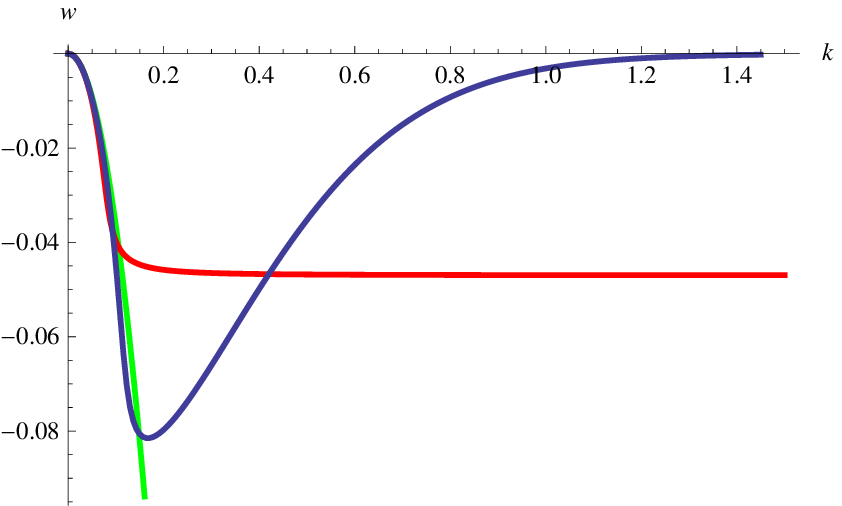}
\end{center}
  \caption{ (Colour online)
The decay width Im$(\ww)$ for $\lgb=-2.5$: The blue line shows the
behaviour of the lowest quasinormal mode calculated numerically. The
red curve corresponds the second-order hydrodynamic approximation
\reef{sound}. The green curve corresponds the next order Taylor
expansion \reef{soundf} arising from the second-order dispersion
relation.}\label{fig45}
\end{figure}

We consider the second-order hydrodynamics and the behaviour of the
lowest sound quasinormal mode in more detail in figures \ref{fig4}
and \ref{fig45}, which show results for $\lgb=-2.5$ -- note that the
latter is outside the ``healthy'' ranges in both \reef{results1} and
\reef{results2}. Using the truncated second-order equations only
yields physically reliable results for $\ka\ll 1$, where the
dispersion relation can be Taylor-expanded as in \reef{soundf}.
These Taylor expansions for the phase velocity and the width, \ie
Re$(\ww)/\ka$ and Im$(\ww)$, keeping only the $O(\ka^2)$ terms are
illustrated with the green curves in the two figures. On the other
hand, the causality analysis of section \ref{second} treats the
dispersion relation \reef{sound} as exact and the results are shown
with the red curves. In this case as $\ka\to\infty$, the phase
velocity rises to Re$(\ww)/\ka=v^{front}_{[\rm sound]}\simeq1.126$
and the width also reaches a finite value asymptotically
\begin{equation}
\lim_{|\ka|\to\infty}\ \frac{\Im(\ww)}{\ka}\bigg|_{\rm
[sound]}=-\frac{1}{\pi\,\tau_\Pi T}\,\frac{\eta/s}{\tau_\Pi
T+4\,\eta/s}\simeq -0.047\,. \eqlabell{wdo}
\end{equation}
As is evident in the figures, these ``exact'' dispersion relations
only match the Taylor expansion \reef{soundf} for small values of
$\ka$.\footnote{In fact, it may seem that, in figures \ref{fig4} and
\ref{fig45}, the two sets of curves begin to separate at
surprisingly small wave numbers (\ie $\ka\sim0.05$ to $0.10$). This
observation can be explained by examining the Taylor series to
higher orders. We find roughly an expansion in
$[(1-4\lgb)^2\ka^2]^n$ for large $\lgb$ and so with $\lgb=-2.5$, the
Taylor series should only be expected to match the ``exact'' result
for $\ka\ll 1/11$.}

To emphasize the limitations of treating \reef{sound} as an
``exact'' dispersion relation, figures \ref{fig4} and \ref{fig45}
also show numerical results for the behavior of the sound mode, \ie
the lowest quasinormal mode with the blue curves. As expected, all
of the different curves agree at small $\ka$ but not at large $\ka$.
While not unexpected, we wish, in particular, to point out that the
striking differences between the actual behaviour of the sound mode
and the second-order hydrodynamic dispersion relation \reef{sound}.
Figure \ref{fig45} shows that the actual width decays rapidly to
zero in contrast to the finite asymptotic limit, given in \reef{wdo}
above, for the second-order dispersion relation. Similarly in figure
\ref{fig4}, the actual phase velocity rises beyond $v^{front}_{[\rm
sound]}\simeq1.126$, the asymptotic limit found for second-order
hydrodynamics,  but then appears to decay back towards one as
$\ka\to\infty$. While the numerical results shown are already
becoming less reliable for $\ka>1$,\footnote{These numerical
difficulties are correlated to the dramatic decrease in
$|\Im(\ww)|$.} it seems that  the limit $\ka\to\infty$ should
produce a front velocity which respects causality.\footnote{Of
course, the higher quasinormal modes are expected to violate
causality with $v_{max}\simeq 1.415$, using the analysis of the
effective Schr\"odinger potential \reef{u0sound}.}

Despite the fact that the results of \reef{sound} may have little
resemblance to the physical behaviour of the sound mode at large
$\ka$, it remains important that the truncated second-order
hydrodynamic equations present a robust mathematical framework in
certain situations. For example, numerical simulations of the
strongly coupled quark-gluon plasma \cite{simulate} implicitly
extrapolate the hydrodynamic equations to the smallest numerical
scales, even though the physics of interest is in the hydrodynamic
regime. In this situation, having a system of hyperbolic equations
which is causal is of course essential.

Our holographic construction with Gauss-Bonnet gravity provides a
simple toy model in which the dual CFT is completely specified by
two central charges $c$ and $a$. The exact relation of these CFT
parameters to the gravitational couplings in the action \reef{gbl}
is given by \cite{two,mam}
 \beqa
c&=&\frac{\pi^2}{2^{3/2}}\,\frac{L^3}{\ell_P^3}\,
(1+\sqrt{1-4\lgb})^{3/2}\,\sqrt{1-4\lgb}\,,\nonumber\\
a&=&\frac{\pi^2}{2^{3/2}}\,\frac{L^3}{\ell_P^3}\,
(1+\sqrt{1-4\lgb})^{3/2}\,\left(3\sqrt{1-4\lgb}-2\right)\,,
 \labell{central}
 \eeqa
and hence
 \beq
\frac{c-a}{c}=2\left(\frac{1}{\sqrt{1-4\lgb}}-1\right)\,.
\eqlabell{defd}
 \eeq
As $\lgb$ alone fixes this last combination, we may re-express our
causality constraints \reef{results2} in terms of the central
charges, which yields the elegant result:
 \beq
 -\frac 12 \le \frac{c-a}{c} \le \frac 12 \ .\labell{nice}
 \eeq
Here we have focused on the fundamental constraint \reef{results2}
rather than the second-order hydrodynamic constraint \reef{results1}
and have found that if the difference in the central charges grows
too large, some linearized fluctuations in the model propagate
faster than the speed of light.

These constraints are also intriguing in comparison to the analysis
of four-dimensional CFT's presented by Hofman and Maldacena
\cite{cft}. They consider ``experiments'' in which the energy flux
was measured in various directions at null infinity after a local
disturbance was created in the stress energy. This energy flux was
found to be controlled by the three-point function of the stress
tensor in the CFT. Further it was shown that the parameters fixing
this three-point function must be constrained in order that the
energy flux was always positive. The holographic dual describing
this CFT ``experiment'' would involve both curvature-squared and
curvature-cubed interactions. However, if the CFT was restricted so
as to eliminate the curvature-cubed interaction in the gravitational
dual,\footnote{This restriction would be realized in any
supersymmetric CFT.} then the positive energy constraint reduced to
a constraint on the central charges, which in fact precisely matches
that given in \reef{nice}. The precise agreement of the upper bound
was already noted in \cite{cft}.

One can go further in that measurements in the above ``experiment''
can be organized into three different channels, just as for the
graviton fluctuations in section \ref{exact}. With this
classification, the upper bound in \reef{nice} comes from the
appearance of negative energy flux in the scalar channel in
\cite{cft} while it arises from causality violation in the same set
of fluctuations in the holographic calculations \cite{vi2a}.
Similarly, the lower bound is set to avoid problematic behaviour in
the sound channel in both approaches. Then we may also note that
while it did not set a fundamental constraint, causality violations
also appear in the shear channel at the critical value given in
\reef{critlgb}: $\lgb^{shear}=-\frac 34$. This result then
translates to a critical value $(c-a)/c|_{shear}=-1$, which again
precisely matches that for the appearance of negative energy fluxes
in the shear channel. Hence, at least within this holographic model,
we have drawn a precise correlation between the appearance of
negative energy fluxes and of superluminal signals in various
channels.

We close with a few comments about other potential instabilities in
this holographic model. It was observed in \cite{vi2} that a new
instability arises in the dual plasma at $\lgb=-1/8$. At this point,
the effective Schr\"odinger potential develops a small well where
$U^0<0$ just in front of the horizon (\ie near $u=1$). For
sufficiently large $\ka$ (and $|\lgb|$), this well will support
unstable quasinormal modes, as described in
\cite{spectre}.\footnote{It was later observed that this instability
seems to be increased by a chemical potential \cite{korea}.}
Examining \reef{u0sound} reveals similar behaviour and hence
instabilities for $\lgb>1/8$ in the sound channel, however, this
problem only appears outside of the range \reef{results2} allowed by
causality. Formally, the effective potential \reef{u0sh} in the
shear channel shows a similar behaviour for $\lgb>1/4$ but this is
again of no consequence since, as noted before, our entire analysis
is only valid in the regime $\lgb<1/4$. Hence the instability in the
scalar channel is the only one that appears within the physical
regime \reef{results2}. This instability does not correspond to a
fundamental pathology with the theory but rather indicates that the
uniform plasma becomes unstable with respect to certain non-uniform
perturbations. It would of course be interesting to follow the full
nonlinear effect of these instabilities. On the gravitational side,
this instability seems similar in certain respects to the
Gregory-Laflamme instability for black strings \cite{gl}.

\section*{Acknowledgments}
After this work was finished, we became aware of an upcoming paper
\cite{hj} which has significant overlap with sections \ref{exact}
and \ref{discuss}. We would like to thank Diego Hofman for sharing
an early draft of \cite{hj} with us. It is also a pleasure to thank
Karl Landsteiner, Paul Romatschke, Aninda Sinha and Andrei Starinets
for useful correspondence and conversations. Research at Perimeter
Institute is supported by the Government of Canada through Industry
Canada and by the Province of Ontario through the Ministry of
Research \& Innovation. AB gratefully acknowledges further support
by an NSERC Discovery grant and support through the Early Researcher
Award program by the Province of Ontario. RCM also acknowledges
support from an NSERC Discovery grant and funding from the Canadian
Institute for Advanced Research.

\appendix
\section{Coefficients of \eqref{qsound}}\label{appa}

\begin{equation}
\begin{split}
&\calc_{sound}^{(1)}=\biggl(\hf u\, (\hf \cala^2 \ka^2 +2 \cala^2
\ka^2-3 \ww^2)-2 \hf^2 u\,
(\hf \cala^2 \ka^2 +10 \cala^2 \ka^2-12 \ww^2) \lgb\\
&-4 \hf^3 u\, (\hf \cala^2 \ka^2 -14 \cala^2 \ka^2+18 \ww^2)
\lgb^2+8 \hf^4 u\,
(\hf \cala^2 \ka^2 -6 \cala^2 \ka^2+12 \ww^2) \lgb^3\\
&-48 \hf^5 u\, \ww^2 \lgb^4\biggr)^{-1} \times
\biggl(-4 \cala^2 \ka^2+6 \ww^2-3 \hf \ww^2 +4 \hf \cala^2 \ka^2 -3 \hf^2 \cala^2 \ka^2 \\
&-2 \hf  (12 \ww^2-3 \hf \ww^2 +8 \hf^2 \cala^2 \ka^2 -4 \cala^2
\ka^2-14 \hf \cala^2 \ka^2 )
\lgb+24 \hf^2  (-4 \cala^2 \ka^2\\
&+\hf \ww^2 +\ww^2+\hf^2 \cala^2 \ka^2 +2 \hf \cala^2 \ka^2 )
\lgb^2-8 \hf^4
(\hf \cala^2 \ka^2 +6 \cala^2 \ka^2+9 \ww^2) \lgb^3\\
&+48 \hf^5 \lgb^4 \ww^2 \biggr)
\end{split}
\eqlabell{csound1}
\end{equation}
\begin{equation}
\begin{split}
&\calc_{sound}^{(2)}=\biggl(\hf^2 u^2\left[-(\hf \cala^2 \ka^2 +2
\cala^2 \ka^2-3 \ww^2) +2 \hf
 (12 \cala^2 \ka^2+2 \hf \cala^2 \ka^2 -15 \ww^2) \lgb\right.\\
&-24 \hf^2 (-5 \ww^2+4 \cala^2 \ka^2) \lgb^2-16 \hf^3
(15 \ww^2-10 \cala^2 \ka^2+\hf \cala^2 \ka^2 ) \lgb^3\\
&\left. +16 \hf^4 (15 \ww^2-6 \cala^2 \ka^2+\hf \cala^2 \ka^2 )
\lgb^4-96 \hf^5 \ww^2 \lgb^5\right]\biggr)^{-1}
\times \biggl(3 \ww^4u-4 \hf \cala^2 \ka^2\\
&+\hf^2 \cala^4 \ka^4u +8 \hf^2 \cala^2 \ka^2-4 \hf^3 \cala^2 \ka^2
-4 \ww^2 \hf \cala^2 \ka^2u
-2 \ww^2 \cala^2 \ka^2u+2 \cala^4 \ka^4 \hf u \\
&+2 \hf (5 \hf \ww^2 u \cala^2 \ka^2-8 \hf^3 \cala^2 \ka^2 -15 \ww^4
u-28 \hf \cala^2 \ka^2
-8 \cala^4 \ka^4 u+8 \cala^2 \ka^2\\
&+2 \hf^2 \cala^4 \ka^4 u+24 \ww^2 \cala^2 \ka^2 u-8 \hf \cala^4
\ka^4 u+28 \hf^2 \cala^2 \ka^2) \lgb
-4 \hf^2 (-30 \ww^4 u\\
&-12 \hf \ww^2 u \cala^2 \ka^2-32 \cala^4 \ka^4 u-4 \hf^2 \cala^2
\ka^2 +2 \hf^2 \cala^4 \ka^4 u
+12 \hf \cala^4 \ka^4 u+40 \hf \cala^2 \ka^2 \\
&+60 \ww^2 \cala^2 \ka^2 u-24 \cala^2 \ka^2-11 \hf^3 \cala^2 \ka^2)
\lgb^2
-8 \hf^3 (2 \hf^2 \cala^4 \ka^4u-56 \ww^2 \cala^2 \ka^2u\\
&+3 \hf^3 \cala^2 \ka^2-24 \cala^4 \ka^4 \hf u+22 \hf^2 \cala^2
\ka^2 +26 \ww^2 \hf \cala^2 \ka^2 u
-24 \hf \cala^2 \ka^2+30 \ww^4u\\
&+24 \cala^4 \ka^4u) \lgb^3+16 \hf^4 (16 \ww^2 \hf \cala^2 \ka^2u
-18 \ww^2 \cala^2 \ka^2u-6 \cala^4 \ka^4 \hf u
+6 \hf^2 \cala^2 \ka^2 \\
&+\hf^2 \cala^4 \ka^4u +15 \ww^4u) \lgb^4-96 \hf^5 u\,\ww^2
(\ww^2+\hf \cala^2 \ka^2 ) \lgb^5\biggr)
\end{split}
\eqlabell{csound2}
\end{equation}
where $\hf$ is given by \eqref{deff}.

\section{Coefficients of \eqref{qshear}}\label{appb}

\begin{equation}
\begin{split}
&\calc_{shear}^{(1)}=\biggl(-\hf u\, (-\ww^2+\hf \cala^2 \ka^2)+4
\hf^2 u\,
 (\hf \cala^2 \ka^2 -2 \ww^2+\ka^2 \cala^2) \lgb\\
&-4 \hf^3 u\, (\hf \cala^2 \ka^2 -6 \ww^2+4 \ka^2 \cala^2) \lgb^2+
16 \hf^4 u\,
(-2 \ww^2+ \cala^2 \ka^2) \lgb^3+16 \hf^5 u\, \ww^2 \lgb^4\biggr)^{-1}\\
& \times \biggl(\hf^2 \cala^2 \ka^2 -2 \ww^2+\hf \ww^2 -2 \hf  (4
\hf \cala^2 \ka^2 +\hf \ww^2
-4 \ww^2) \lgb+8 \hf^2  (-\hf \ww^2 -\ww^2 \\
&+2 \ka^2 \cala^2) \lgb^2+24 \hf^4 \lgb^3 \ww^2 -16 \hf^5 \lgb^4
\ww^2 \biggr)
\end{split}
\eqlabell{cshear1}
\end{equation}
\begin{equation}
\begin{split}
&\calc_{shear}^{(2)}=\biggl((2 \lgb \hf -1)^2 \hf^2 u\biggr)^{-1}
\times \biggl(\ww^2-\hf \cala^2 \ka^2 +4 \hf  (\ka^2 \cala^2-\ww^2)
\lgb +4 \hf^2 \lgb^2 \ww^2 \biggr)
\end{split}
\eqlabell{cshear2}
\end{equation}
where $\hf$ is given by \eqref{deff}.

\end{document}